\documentclass[12pt,a4paper,fleqn,notitlepage,onecolumn]{article}

\usepackage{txfonts}
\usepackage{graphicx}
\usepackage{natbib}

\usepackage{geometry}
\usepackage{fancyhdr} 
\newcommand{\arxiv}{arXiv}

\newcommand{\maxi}{MAXI\,J0556-332}

\title{Monte Carlo error analyses of Spearman's rank test}

\author{Peter~A.~Curran\footnote{email: {\tt peter.curran@curtin.edu.au}}\\
\footnotesize{International Centre for Radio Astronomy Research,}\\
\footnotesize{Curtin University, GPO Box U1987, Perth, WA 6845, Australia}\\
}
\date{}

\begin{document}

\maketitle


\begin{center}
\begin{minipage}{15cm}
\small{Spearman's rank correlation test is commonly used in astronomy
  to discern whether a set of two variables are correlated or
  not. Unlike most other quantities quoted in astronomical literature,
  the Spearman's rank correlation coefficient is generally quoted with
  no attempt to estimate the errors on its value. This is a practice
  that would not be accepted for those other quantities, as it is
  often regarded that an estimate of a quantity without an estimate of
  its associated uncertainties is meaningless. This manuscript
  describes a number of easily implemented, Monte Carlo based methods
  to estimate the uncertainty on the Spearman's rank correlation
  coefficient, or more precisely to estimate its probability
  distribution.
}
\end{minipage}
\end{center}

\section{Introduction}\label{section:intro}

Spearman's rank correlation test \citep{Spearman1904:AJP15} is
commonly used in astronomy to discern whether a pair of two variables
are correlated or not. It has an advantage over the Pearson
correlation test\footnote{Here error analyses of the Spearman test are
  discussed, though the methods can (and should) also be applied to
  the Pearson test.} -- which requires a linear relationship between
the two variables -- as it is non-parametric. This non-parametric
nature of the test is useful as it is model independent so long as the
relationship is monotonic, i.e. is a steadily increasing or decreasing
function.  Unlike other quantities in literature, Spearman's rank
correlation coefficient, $\rho$, is generally quoted with no attempt
to estimate the uncertainties on its value; though the data used to
calculate the coefficient will almost certainly have some measurement
uncertainty, in at least one of the variables.

In the fields of e.g. bio-medicine and clinical medicine, where
Spearman's rank test is regularly used, the method of Monte Carlo
\emph{bootstrapping} \citep{Efron1979:AS7} is used to estimate the
confidence intervals of the coefficient (e.g.,
\citealt{Haukoos2005:AEM.12}); this involves resampling the data by
drawing random entries from the original data set to create multiple
resampled data sets of the same size as the original. The method does
have limitations, as described by \cite{Haukoos2005:AEM.12}, primarily
that the method assumes that the sample is representative of the
overall population which is of particular concern for small sample
sizes. However, few other methods are available to estimate the
uncertainties.

The principal difference between astronomical and clinical data is
that the astronomical data will often, though not always, have
associated measurement uncertainties (e.g., flux, magnitude, distance)
while clinical data is often count-based (e.g., frequency of success
in clinical trials) with no intrinsic probability distribution on the
individual data points. Of course count based astronomical data is
commonplace (e.g. population studies) but even in these cases an
attempt is often made to approximate the probability distribution of
the data by assuming e.g. Poissonian uncertainties.  Here three Monte
Carlo methods of error analyses of the Spearman's rank coefficient --
the bootstrap/resampling method, the perturbation method and composite
method -- are described and demonstrated using a sample data set.
Importantly, the latter two methods exploit the uncertainties or
probability distributions which are generally associated with
astronomical data.

\section{Method}\label{section:method}

Given a data set consisting of $N$ data pairs, $X_i$ and $Y_i$, each
individual entry is assigned an ascending order rank, $RX_i$ and
$RY_i$, and the Spearman's rank correlation coefficient, $\rho$, for
the sample is calculated from the square of the difference of the two
ranks for each pair by 
\[ \rho = 1 - \frac{6\displaystyle\sum_{i=1}^{N} (RX_i - RY_i)^2}{N(N^2 -1)} . \]
A coefficient of 0 corresponds to no correlation between the
variables, while a value of $+1$ or $-1$ corresponds to a perfect
increasing or decreasing monotonic correlation. The significance of
the correlation may be calculated in a number of ways, such as a
Student's t-test \citep{Zar1972:JASA67}. Here the z-score, $z$, is
used since it approximately follows a Gaussian, or normal,
distribution in this case, i.e., a z-score of $z$ approximately
corresponds to a Gaussian significance of the correlation of $\sigma
\approx z$.  The z-score is calculated by $z = F(\rho)
\sqrt{\frac{N-3}{1.06}}$, where $F(\rho) = arctanh(\rho) =
\ln[(1+\rho)/(1-\rho)]/2$ is the Fisher transformation of the
correlation coefficient, which goes to infinity as $\rho$ goes to
1. 

The aforementioned bootstrap or resampling method of error analysis
involves creating $M$ new data sets, each consisting of $N$ data
pairs, $x_i$ and $y_i$, where for statistical significance, $M \gtrsim
1000$. Each of these new pairs is a randomly chosen pair from the
original data set, $X_j$ and $Y_j$, where $j$ is the randomly chosen
entry, e.g., $x_i = X_j$ and $y_i = Y_j$, such that some of the
original pairs may appear more than once in a given data set or not at
all.  The data points are again assigned a rank, $Rx_i$ and $Ry_i$,
and the Spearman's rank correlation coefficient and z-score for each
of these $M$ new data sets calculated. The distributions of the
returned values are used to estimate the probability distribution of
the two quantities, by normalising so that the integral is equal to
one.  In the simplest case of this probability distribution being
Gaussian, the estimate of the correlation coefficient, $\hat{\rho} =
\bar{\rho}$, the average of the calculated values and the estimate of
the error is the Gaussian width of the distribution, $\sigma_{\rho}$,
i.e., the standard deviation of the calculated values; the z-score is
similarly estimated as $\bar{z} \pm \sigma_{z}$\footnote{For a much
  fuller discussion of using probability distributions, Gaussian or
  otherwise, to estimate quantities see \cite{Andrae2010:arXiv1009},
  which also includes discussion on the use of Monte Carlo methods to
  estimate errors.}.

The resampling method obviously does not take into account the
uncertainties, $\Delta X_i$ and $\Delta Y_i$, likely associated with
the pairs. To do so a perturbation method may be used, where one
creates multiple new data sets ($M \gtrsim 1000$) consisting of $N$
data pairs, $x_i$ and $y_i$, each of which is perturbed from the
original values, $X_i$ and $Y_i$, by adding a random
Gaussian\footnote{Here, as is often done, it is assumed that the
  measurement uncertainties are Gaussian; however, these methods only
  require that the probability distribution of the measurement
  uncertainties are known.}  number times the uncertainty on that
point, e.g., $x_i = X_i + \mathcal{G} \times \Delta X_i$, where
$\mathcal{G}$ is a number, drawn randomly from a Gaussian distribution
of width 1 and centered on 0.  This is done independently per point
(assuming that the uncertainties are independent) so that a different
random Gaussian number, $\mathcal{G}$, is used to perturb the $X$ and
$Y$ values. The Spearman's rank correlation coefficient and z-score are
then calculated for each of these $M$ data sets and the distribution
of returned values used as an estimate of the probability distribution
of the two quantities, as before.

Alternatively, the composite method may be used, combining the
traditional resampling method and the above outlined perturbation
method. In this case each new data pair, $x_i$ and $y_i$, is perturbed
from a random entry of the original data set, $X_j$ and $Y_j$, where
$j$ is the randomly chosen entry, e.g., $x_i = X_j + \mathcal{G}
\times \Delta X_j$. Again, the random Gaussian numbers, $\mathcal{G}$,
must be independent, in so far as that is possible given whatever
random number generator is being used, but the two variables must have
originated from the same source pair. The probability distribution of
the two quantities, $\rho$ and $z$, are again estimated as above.


\section{Example}\label{section:example}

 \begin{figure} 
  \centering 
  \includegraphics[angle=-90,width=8cm]{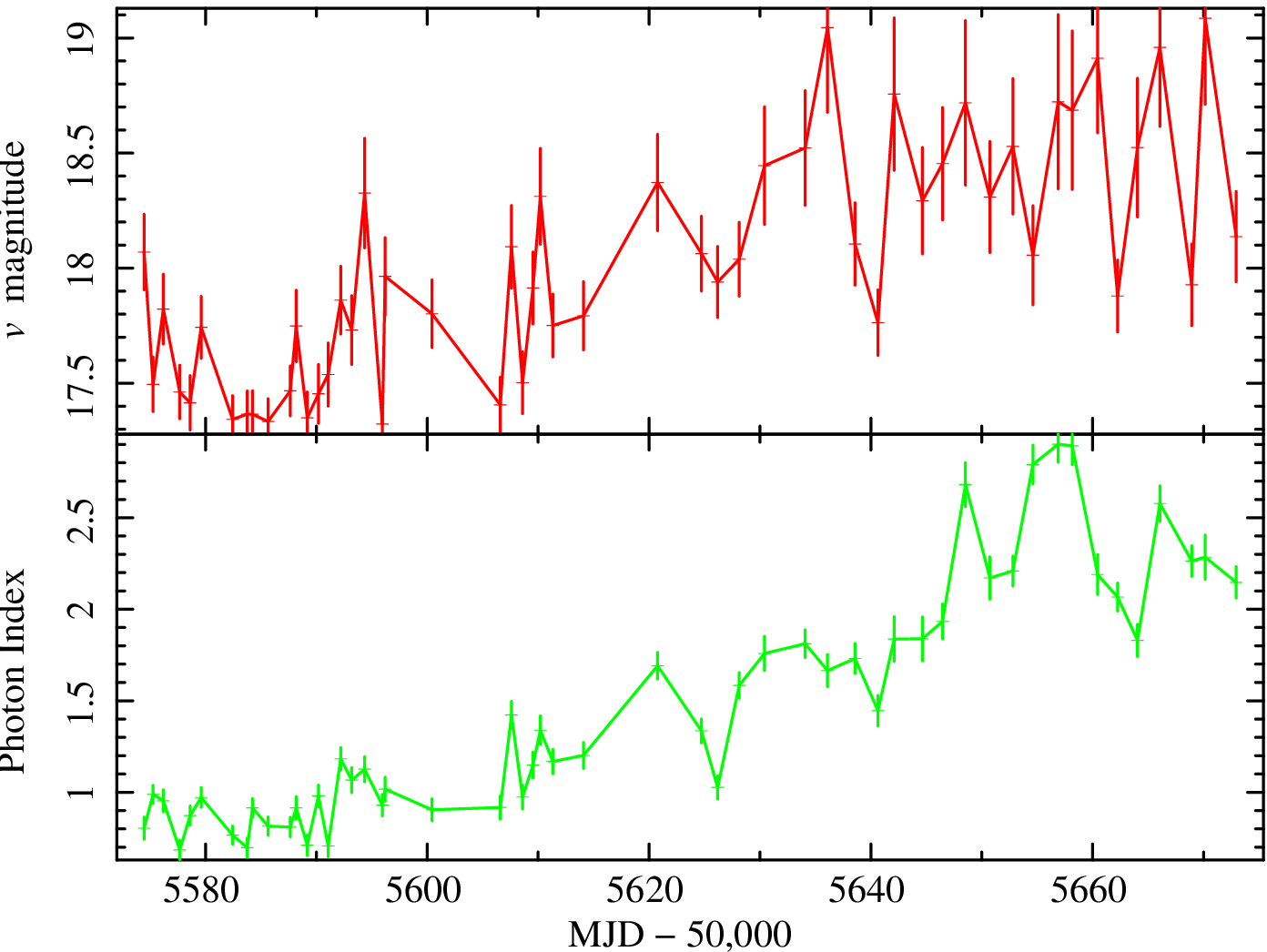}\hspace{3mm}
  \includegraphics[angle=-90,width=8cm]{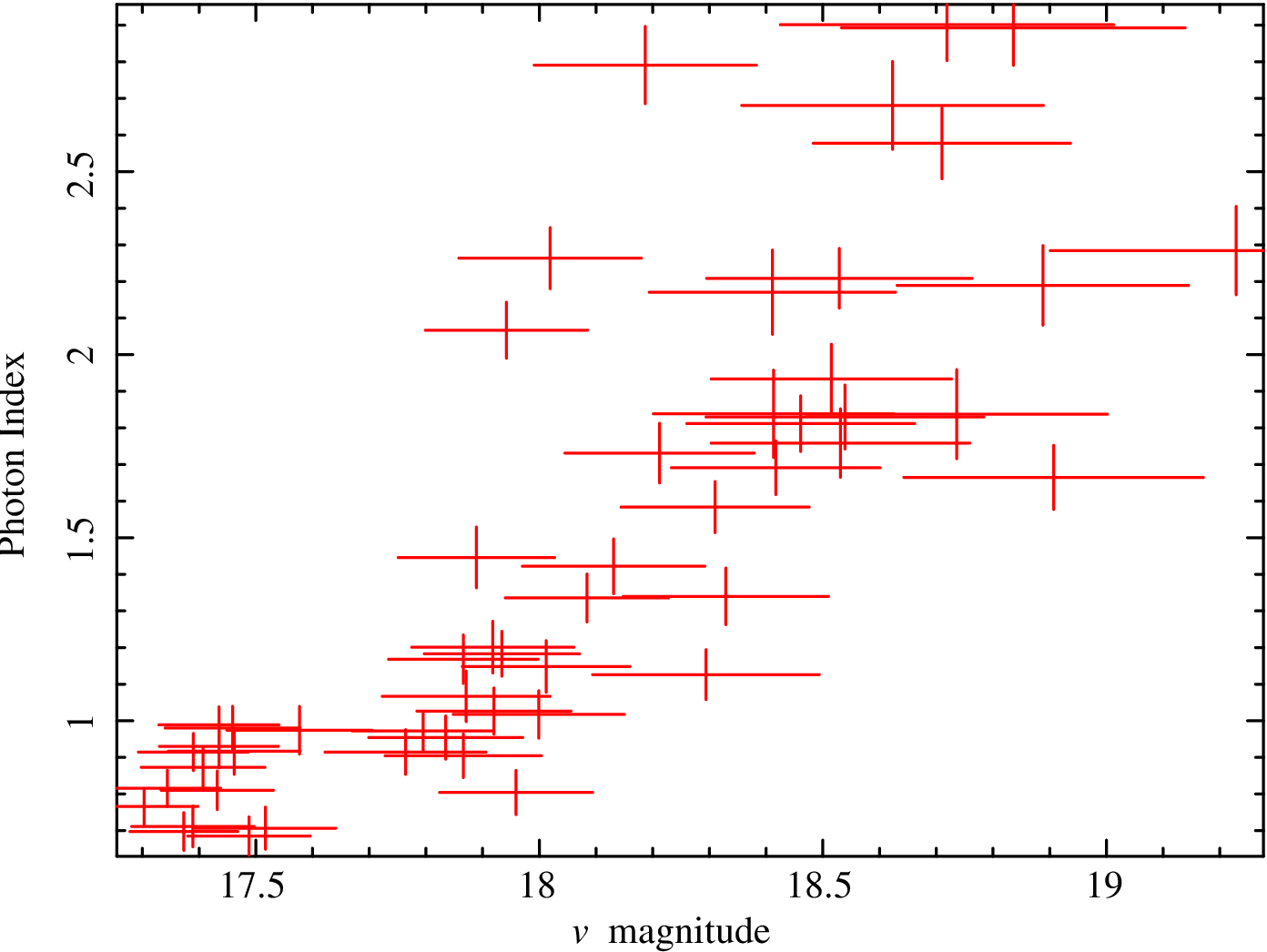}
  \caption{\emph{Left}: MAXI\,J0556-332 $v$-band \emph{Swift}-UVOT
    light curve and photon index of spectral fit of the UVOT
    data. \emph{Right}: the scatter plot of the same data (53 data
    pairs). }
  \label{fig.data} 
\end{figure}

\begin{figure}
  \centering 
  \includegraphics[angle=-90,width=8cm]{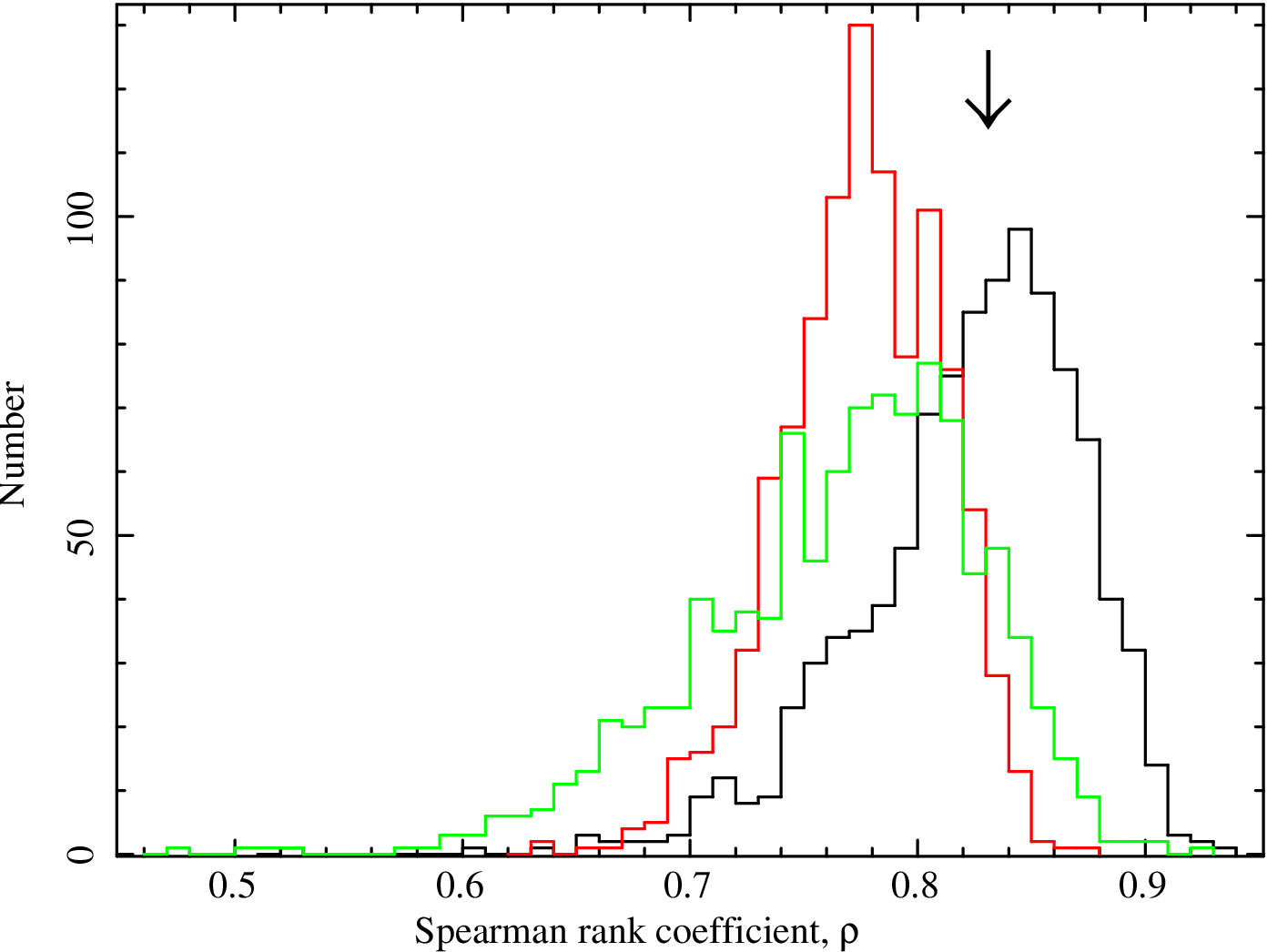}\hspace{3mm}
  \includegraphics[angle=-90,width=8cm]{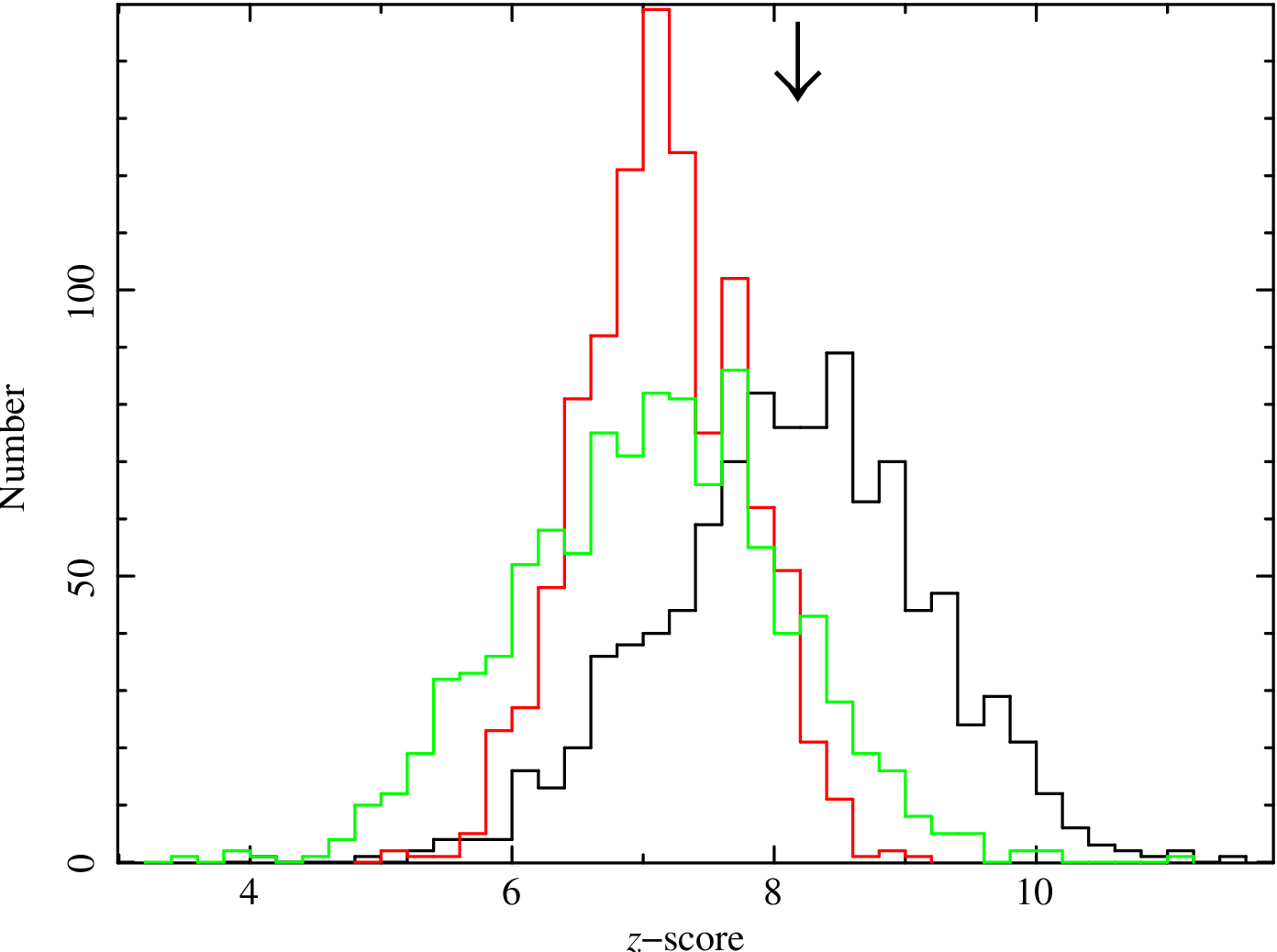}
  \caption{\emph{Left}: Histogram plot of Spearman rank coefficients
    for the MAXI\,J0556-332 data as derived from the resampling method
    (black), the perturbation method (red) and the composite method
    (green).  \emph{Right}: the distribution of the z-scores for those
    values (1000 iterations per test).  In both plots, the arrow
    represents the result of the standard method. }
  \label{fig.spearman} 
\end{figure}

Here I apply the standard method of estimating the Spearman's rank
correlation coefficient (no error) and the different Monte Carlo
methods (resampling, perturbation and composite)\footnote{For these
  results, the methods were implemented within a C program, \emph{MCSpearman}
  (available online; \citealt{Curran:2015ascl}), but are easily implemented in many
  languages or programs.} to real physical data from the X-ray
transient, \maxi. During its 2011 outburst, \maxi\ was observed by the
UVOT instrument on board the \emph{Swift} satellite. When the photon
indices of these optical observations were compared to the $v$-band
magnitudes, an apparent correlation was clear
(Figure\,\ref{fig.data}). In fact, the Spearman's rank correlation
coefficient was $0.83$, implying a z-score significance of $z = 8.2$,
corresponding to a significance of the correlation of $\approx
8.2\sigma$ ($N=53$). If one were to take the basic step of applying
the resampling method to this ($M=1000$ in this and the following
cases), which is not often done, the distributions of $\rho$ and $z$
plotted in Figure\,\ref{fig.spearman} (black) are obtained. Assuming
that these may be described as Gaussian distributions (clearly not the
case for the correlation coefficient but a reasonable simplifying
assumption nonetheless), their averages and standard deviations are as
given in Table\,\ref{table:results}.

It is clear from Figure \ref{fig.data} that there is significant
uncertainties on both the spectral indices and the magnitudes which
may reduce the significance of the correlation, and should be taken
into account. Applying the perturbation method one finds that, indeed,
the distribution and average of the correlation coefficient is
reduced, as is the z-score significance (Figure\,\ref{fig.spearman},
red).  Applying the composite method, one finds that it returns
similar values to the perturbation method but with a wider
distribution of values (Figure\,\ref{fig.spearman}, green). In fact,
the plotted distributions of the composite method clearly demonstrate
that what was considered a correlation at the $\approx 8.2\sigma$
level is better described by a significance of $\approx (7.1 \pm
1.0)\,\sigma$, which has a non-negligible probability of being
$<5\sigma$.

\begin{table}	
  \centering	
  \caption{Average $\pm$ standard deviations of the correlation
    coefficients, $\rho$, and z-scores, $z$, for the different methods
    as applied to the MAXI\,J0556-332 data. }
  \label{table:results} 	
  \begin{tabular}{l l l} 
    \hline\hline
    method & $\rho$ & $z$  \\ 
    \hline 
    standard     &  0.83 & 8.2 \\
    bootstrap    & $0.82 \pm 0.05$  & $8.2 \pm 1.0 $ \\
    perturbation & $0.78 \pm 0.04$  & $7.2 \pm 0.6 $ \\
    composite    & $0.77 \pm 0.06$  & $7.1 \pm 1.0 $ \\
    \hline 
  \end{tabular}
\end{table}


\section{Discussion}\label{section:discussion}

Only one test case of these methods is presented so general
conclusions should not be drawn without serious caution. For example,
the results will be heavily dependent on the number of data points,
the distribution of these data points in $x$ and $y$, and, in the
cases of the perturbation and composite methods, the size of the
uncertainties on the data points. For a discussion regarding the
number and distribution of data points see \citet{Haukoos2005:AEM.12}.

Clearly, taking the data uncertainties into account weakens the
correlation between the two variables; as the data uncertainties go to
zero, the perturbation method will tend to a delta function at the
value returned by the standard method, while the results of the
composite method will tend to those of the resampling method.

It is important to understand the difference between the two
(non-composite) methods and the distribution they return.  The
resampling method estimates the uncertainty of the correlation
coefficient given the uncertainty of the sample, i.e., the sample
being tested is only a sub-sample of the population of all possible
data pairs and the resampling method estimates the uncertainties
associated with the lack of information of all those data pairs.  The
perturbation method estimates the uncertainty of the correlation
coefficient, given only the uncertainties on the data points, i.e.,
this method assumes the given sample is absolutely representative of
the population, or alternatively, the method only estimates the error
of the given pairs in the sample, not the population as a whole.
In some circumstances, one may only want to estimate the correlation
coefficient (and uncertainty) of the given sample in which case the
perturbation method should be used. In other circumstances, it is the
uncertainty associated with the population, including its unknown
entries, which dominates, and the resampling method should be used. In
many cases, though only one of these dominate the probability
distribution of the correlation coefficient, both should be taken into
account via the composite approach.

\section{Conclusion}\label{section:conclusions}

When calculating the Spearman's rank correlation coefficient of a data
set, an estimate of the uncertainty of the coefficient is required,
just as it is for most other quantities in astronomy. Furthermore,
when the given data set has uncertainties on the individual entries,
these uncertainties must be taken into account when estimating the
correlation coefficient. Here I have suggested and discussed three
Monte Carlo methods of accounting for the uncertainties in the data
points which are easily implemented and return significant information
regarding the probability distribution of the correlation coefficient.


\paragraph*{Acknowledgements}
The author thanks Anna D. Kapi\'{n}ska for constructive comments and
suggestions on drafts of this manuscript.  The work was supported by
the Centre National d'Etudes Spatiales (CNES), France through MINE:
the Multi-wavelength INTEGRAL NEtwork.

\end{document}